# ALPC IS IN DANGER: ALPCHECKER DETECTS SPOOFING AND BLINDING

Anastasiia Kropova
Bachelor of Information Security, MEPhI
Moscow, Russia
kropovaanastasiia@gmail.com

Igor Korkin, PhD
Security Researcher
Moscow, Russia
igor.korkin@gmail.com

**ABSTRACT**

The purpose of this study is to evaluate the possibility of implementing an attack on ALPC connection in the Windows operating system through the kernel without closing the connection covertly from programs and the operating system and to propose a method of protection against this type of attacks. Asynchronous Local Procedure Call technology (ALPC) is used in various Windows information protection systems, including antivirus systems (AV) and Endpoint Detection and Response systems (EDR). To ensure the concealment of malicious software, attackers need to disrupt the operation of AV, EDR tools, which in turn can be achieved by destructive impact on the components of the ALPC technology. Examples of such attacks already exist and are covered in this paper. To counteract such new threats, it is necessary to advance the improvement of information security systems and the ALPC security research was conducted. The most difficult case, Windows kernel driver attack, was considered. Three attacks on the ALPC connection were carried out, based on changing the ALPC structures in the kernel memory, which led to creation of illegitimate connections in the system and the disruption of correct connections. ALPChecker protection tool has been developed. The tool was successfully tested on three demonstrated attacks.

**Keywords**: kernel driver, ALPC mechanism, interprocess communication, Connection Port, client process, server process.

## 1. INTRODUCTION

According to statistics Statcounter.com (Statcounter.com, 2023), the most popular operating system for computers is Windows. It is installed on 74.14% of computers. OS X takes the 2nd place – 15.33%, followed by Linux – 2.91%.

With each version, more and more security features appear and improve in Windows, but the system is still not fully protected (Korkin, 2022). Attackers still continue to find gaps in the protection of the operating system, bypassing security features (Klein, Kotler, 2019). The popularity of Windows makes it a desirable target for attackers (Pei et al., 2016).

A huge number of attacks on Windows are carried out in kernel mode using drivers. All loaded kernel drivers use the same address space as the Windows kernel itself. According to Pogonin (Pogonin, Korkin, 2022), there is no built-in solution to control and restrict all memory access between these drivers and sensitive kernel data. Therefore, the Windows operating system continues to be at risk of attacks from kernel drivers.

### Windows security mechanisms cannot prevent driver-based attacks

Various mechanisms have been created in order to protect Windows from attacks in kernel mode. Among them are DSE (Driver Signature Enforcement), prohibiting the download of malicious drivers, PatchGuard in Windows 8.1 and MmProtectDriverSection in Windows 11, protecting DSE from changing the nt!g_CiEnabled kernel variable and disabling it (Pogonin, Korkin, 2022).

Attackers began to use official, Microsoft-signed drivers with known vulnerabilities, embedding files intended for the attacked system in them to bypass security tools. This technique is called BYOVD (Bring Your Own Vulnerable Driver) (MITRE, 2021).

Tsarfati (Tsarfati, 2023) discovered multiple bugs in OEM vendors for peripheral devices and managed to exploit vulnerability in WinIO driver.

The ASEC (Sanseo, 2023) analysis team discovered a Sliver backdoor being installed through what is presumed to be vulnerability exploitation and used the BYOVD malware to incapacitate security products and install reverse shells.

A new hacking campaign exploiting Sunlogin flaws to deploy the Sliver post-exploitation toolkit and launching Windows BYOVD attacks to disable security software was described by Toulas (Toulas, 2023)

Poslušný from ESET in his paper (Poslušný, 2022) describes the most common types of vulnerabilities and provides examples of vulnerable drivers and malicious software exploiting these vulnerabilities.

Among the described programs, the following can be highlighted:

Slingshot cyberespionage platform, which implements its main module as a kernel driver, using signed driver loaders Goad, SpeedFan, Sandra, and ElbyCDIO;

InviziMole package using MS driver vulnerability speedfan.sys to download your malicious driver (Poslušný, 2022);

RobinHood malware using GIGABYTE motherboard driver GDRV.SYS (SecureAuth, 2020) to disable the DSC and install its own driver.

CrowdStrike researchers (Iacob, Ionita, 2022) described how the vulnerable VBoxDrv driver is used to install unsigned ElRawDiskDriver. The driver is used to transfer actions from user mode into kernel mode. Similar EPMNTDRV driver was also used for malicious purposes, malware expoited it to wipe MBR, MFT and files on behalf of the legitimate driver.

In 2018, the first rootkit for UEFI systems was discovered – LoJax. A kernel driver was used to access the system settings, RwDrv.sys and the RWEverything utility, which allows you to quickly read data from memory. The system settings are read into a text file and uploaded back to the system along with the added malicious module (ESET, 2018).

In 2022, Kaspersky experts discovered the UEFI rootkit MoonBounce capable of introducing malicious drivers into the Windows kernel (Lechtik, Berdnikov, Legezo 2022).

Baines (Baines, 2021) form Rapid7 cited in his article about thirty examples of using vulnerable official drivers to commit attacks on the operating system.

Hfiref0x introduced tools such as the driver loader TDL (Hfiref0x, 2019a) to bypass DSE, UPGDSED (Hfiref0x, 2019b) tool, disabling DSE and PatchGuard, kernel driver utility KDU (Hfiref0x, 2022), using vulnerable drivers to access the system.

Magdy and Zohdy (Magdy, Zohdy, 2023) from TrendMicro examined how windows kernel threats evolved before and after the appearance of KMCS kernel mode code signing.

TrendMicro analysts (Magdy, Zohdy, 2022a) categorized kernel-level threats into three clusters based on observable techniques: threats that bypass KMCS, threats, that comply KMCS, using create-your-oen-driver techniques and threats that shift to a lower abstraction level. Researchers analyzed the statistics how these threats affected Windows for the past eight years. The full analysis of more than 60 low-level threats to the windows kernel, observed from 2015 to 2022 can be found in their other article (Magdy, Zohdy, 2022b). It is showed how these threat actors adapts to the current defense mechanisms and how they are evolving their techniques.

TrendMicro researchers (Zahravi, Girnus, 2023) found a campaign that uses a fake employment pretext to install the Enogma Stealer application and steal cryptocurrency information. The attacker exploits CVE-2015-2291, an Intel driver vulnerability.

Arghire (Arghire, 2023) from Securityweek reported that cybercrime group tracked as Scattered Spider has been observed exploiting an old vulnerability in an Intel Ethernet diagnostics driver for Windows in recent attacks on telecom and BPO firms.

Malicious drivers can be signed with revoked and stolen certificates. For example, such a certificate was used by the DirtyMoe rootkit driver (Chlumecký, 2021).

Revoked and stolen certificates can be bought by the criminals on the Dark Web which is one of the main platforms for selling codesigning certificates (Barysevich, 2018). For example, Nvidia (Abrams, 2022) and Frostburn Studios (Voronovitch, 2022) digital certificates were stolen and used to sign malicious drivers.

The paper (Gupta et al.) shows that Windows kernel driver includes bugs that can be easily found by security experts. A lightweight framework POPKORN that harnessed the power of taint analysis and targeted symbolic execution to automatically find security bugs in Windows kernel drivers at scale was introduced. When run against 212 unique Windows drivers, POPKORN reported 38 high impact bugs in 27 unique drivers, with manual verification revealing no false positives. Among the found bugs 31 were previously unknown vulnerabilities that potentially allow for Elevation of Privilege (EoP).

## ALPC mechanism in Windows OS

In this paper, we study a new kernel drivers attack vector to the ALPC mechanism of client-server interaction of Windows, explore its architecture and basic structures. Attacks on ALPC interaction using kernel drivers have been carried out and a new tool for protecting against attacks of this type has been proposed.

Advanced/Asynchronous Local Procedure Call (ALPC) — an advanced system for calling local procedures — the internal mechanism of Interprocess Communication or IPC in Windows. It transfers messages between the client and server process on the same computer, see Figure 1.

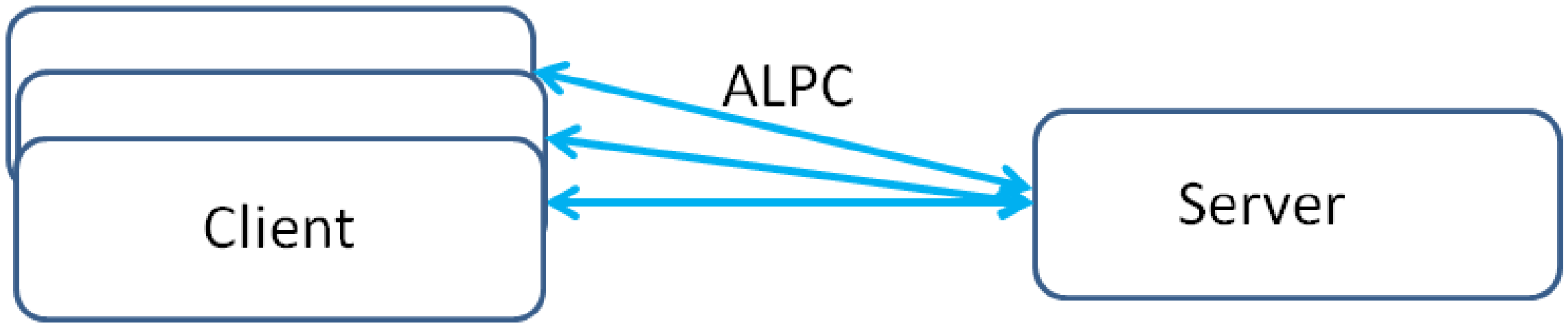

Figure 1. ALPC connection

ALPC is used everywhere due to its scalability, high speed and the ability to send messages of any length. ALPC is used in kernel mode RPC calls (Remote Procedure Call), which are used by the network stack, as a standard instrument of communication for client and server processes over the network. It is used for communication by system processes such as Session Manager (SMSS), Local Session Manager (LSM). The Error Reporting System (WER) receives information about emergency processes via ALPC. Thanks to the use of ALPC by the User Mode Driver Environment (UMDF), drivers can exchange information among themselves.

The User mode monitor and Power Manager use ALPS to communicate with the kernel mode power Manager (for example, when the screen brightness changes).

ALPC is used every time a Windows process or thread is started and during any Windows subsystem operation to communicate with CSRSS (Client Server Run-Time Subsystem). The login program and the Link Security Monitor use ALPC to communicate with LSASS (Local Security Authority Subsystem Service).

The above allows us to conclude that ALPC security threats are threats to the security of all Windows software and the security tools themselves.

The purpose of this part is to raise the issue of ALPC security, analyze the work of ALPC and review existing attacks on ALPC.

To detect driver-based kernel-mode attacks on ALPC the ALPChecker tool will be presented.

ALPChecker prevents spoofing and blinding attacks on ALPC connection by verification of client and server information compliance.

The key feature of ALPChecker allows it to collect and verify kernel data from user-mode without stopping the system. ALPChecker examines all the user and system connections and displays the message about the results of its work.

The remainder of the paper is as follows.

Section 2 provides the description of the ALPC internal structure. The analysis of the ALPC architecture and structures in the kernel memory is given. Examples of port enumeration programs are demonstrated.

Section 3 presents the examples of recent research papers on attacks on ALPC.

Section 4 contains the description of our new spoofing and blinding attacks on ALPC-connections and their results.

Section 5 presents ALPChecker, demonstrates and describes its work.

## 2. THE INTERNAL STRUCTURE OF THE ALPC

This section describes the internal structure of the ALPC, its background, purpose and capabilities. ALPC architecture and structure analysis are presented in this section.

### 2.1. Purpose and capabilities of ALPC

The purpose of the ALPC mechanism is the transmission of messages between the server process and its client processes. An ALPC connection can be established between a kernel-mode component and user-mode processes, or between user-mode processes (Allievi, Ionescu, Russinovich, Solomon 2021).

ALPC was added in the Windows Vista operating system as a replacement for the outdated LPC mechanism that was supplied with the first designs of the Windows NT kernel. The local procedure call, LPC, was a synchronous mechanism of interprocess communication. Clients and servers had to wait for the message to be sent and the appropriate actions to be performed before the applications could be continued. To fix this main drawback, which significantly slowed down the system, an ALPC mechanism for asynchronous calling of local procedures was created. Starting with Windows 7, the LPC in the NT kernel was completely replaced by the ALPC. The vulnerabilities of the LPC, which is emulated at the upper levels of the ALPC to ensure compatibility with all applications, have been eliminated in the ALPC.

ALPC is used for every interaction with a COM object. COM (Component Object Model) technology is widely used in Windows OS – a way of sharing objects and functions inside and outside the process. A variation of this technology DCOM (Distributed COM) allows COM objects to be accessed outside the process, interact over the network and provides access to methods and calls of these objects. As far as DCOM objects are used everywhere in Windows, including applications from the Windows Store, user mode drivers, date and time controls, ALPC is used by almost every program.

When the ALPC mechanism was added to the Windows operating system instead of LPC, it brought stability, high speed of interprocess communication and replaced the Named Pipes method in Windows.

Even a simple program on Windows will have an ALPC connection with at least one process (Ionescu, 2014). All applications and components that previously used well-known Named Pipes, including RPC and DCOM technology using it, now are using undocumented ALPC.

For example, the Notepad application has server connections with svchost.exe (the main process of services loaded from dynamic libraries) and ctfmon.exe (the process associated with the CTF loader, Collaborative Translation Framework, used for handwriting and speech recognition) and client connections to csrss.exe, svchost.exe, lsass.exe and ctfmon.exe (Figure 2).

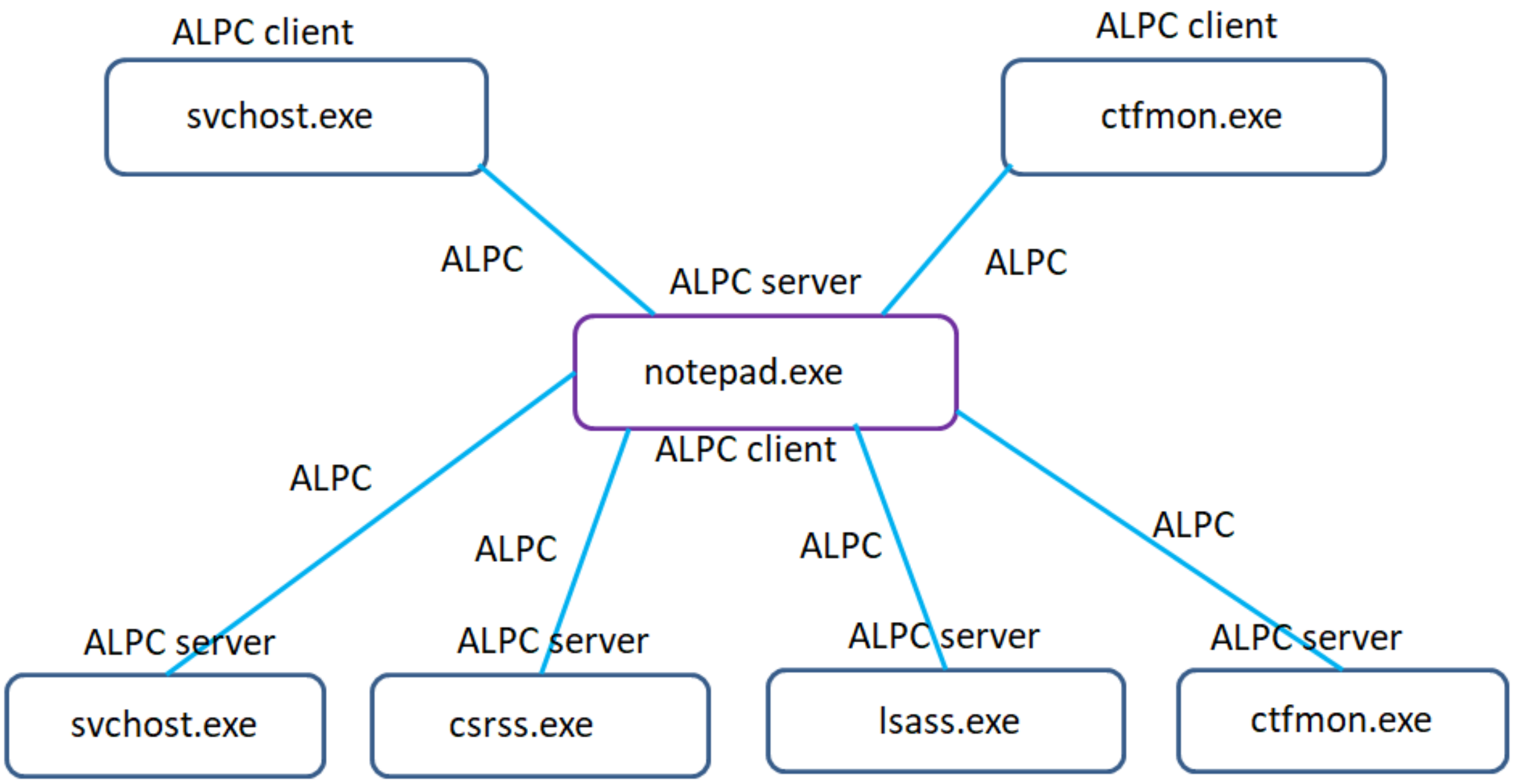

Figure 2. ALPC connections of the notepad.exe application

**Linux analogue**

The closest analogue of ALPC in the Linux operating system is Dbus or Desktop Bus. Desktop Bus is a system that allows applications and services, as well as users and applications, to interact with each other. DBus consists of a control daemon, a console client, and an API for programming languages through which applications can interact with the system (Freedesktop, 2022).

## 2.2 ALPC architecture

The primary components of the ALPC interaction are the ALPC ports. There are four different types of ALPC ports:

• Client Communication Port - the port that the client process uses to communicate with the server, an unnamed port;

• Server Communication Port – the unnamed port port that the server uses to communicate with the client. The server has one communication port for each of its active clients;

• Server Connection Port - the port that is specified in the request to connect to the server, this is a named port. By connecting to this port, clients can connect to the server;

• Unconnected Communication Port – a port that can be used by the client for local communication with itself, an unnamed port. This model was abolished in the move from the LPC to ALPC.

The message management algorithm in ALPC has not changed compared to LPC and uses LPC functions, these functions start with "Lpcp". However, the ALPC functions using them are smaller and more convenient. New kernel-level ALPC functions start with "NtAlpc" and are available from the library ntdll.dll (Garnier, 2008).

In addition to forwarding messages between the server and the client, there is another data transfer

mechanism – a shared partition. Part of the memory is allocated for shared use by the client and the server, and a sequential and identical view of this memory is opened for them. Thus, as much data can be transferred at a time as will fit into a given memory section.

The partition objects and the ALPC itself provide protection against privilege escalation attacks by adding new links to messages, local object mappings, and security modes. Detailed scheme of ALPC interaction can be found in the book (Allievi, 2021).

### 2.3. Analysis and description of ALPC structures in kernel memory

ALPC ports are stored in the operating system memory as the ALPC_PORT structure. This structure contains a set of fields with complete information about all objects of this interaction: the list of ports (PortListEntry), the port at the other end (CommunicationInfo), message queues (mainQueue, PendingQueue, WaitQueue, CanceledQueue, LargeMessageQueue), the owner process of this (OwnerProcess) and others, see Figure 3.

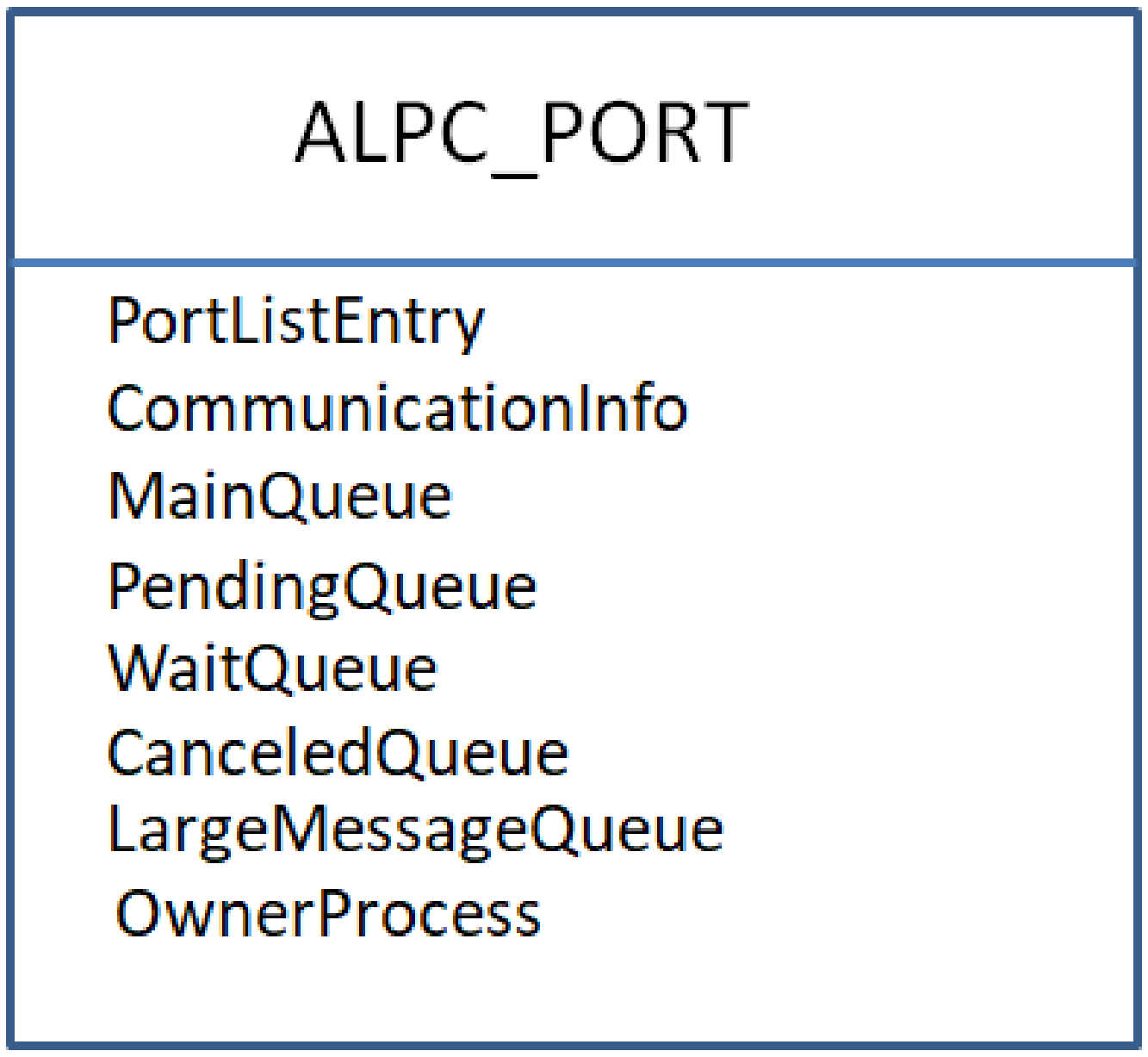

Figure 3. Fields of the ALPC_PORT structure

The CommunicationInfo field, located at the address of the ALPC port with an offset of +0x010, is of the greatest interest for our study. This field is an ALPC_COMMUNICATION_INFO structure. It is this structure that contains information about the connections of this port. The structure contains the fields ConnectionPort - the address of the Connection Port, ServerCommunicationPort – the address of the Server Communication Port and ClientCommunicationPort – the address of the Client Communication Port. Below is a pointer to the connection descriptor table - HandleTable and the connection closure message – CloseMessage, see Figure 4.

The structure of the message being sent, KALPC_MESSAGE, was also considered, see Figure 5.

This structure contains information about the connection, the port queue – all the opened ALPC ports in the system, the port that sent the message, message attributes, threads and buffers.

Oxcsander (Oxcsandker, 2022) conducted a dynamic analysis of the ALPC structures, ALPC message flow. His work provides a detailed scheme for creating ALPC connection.

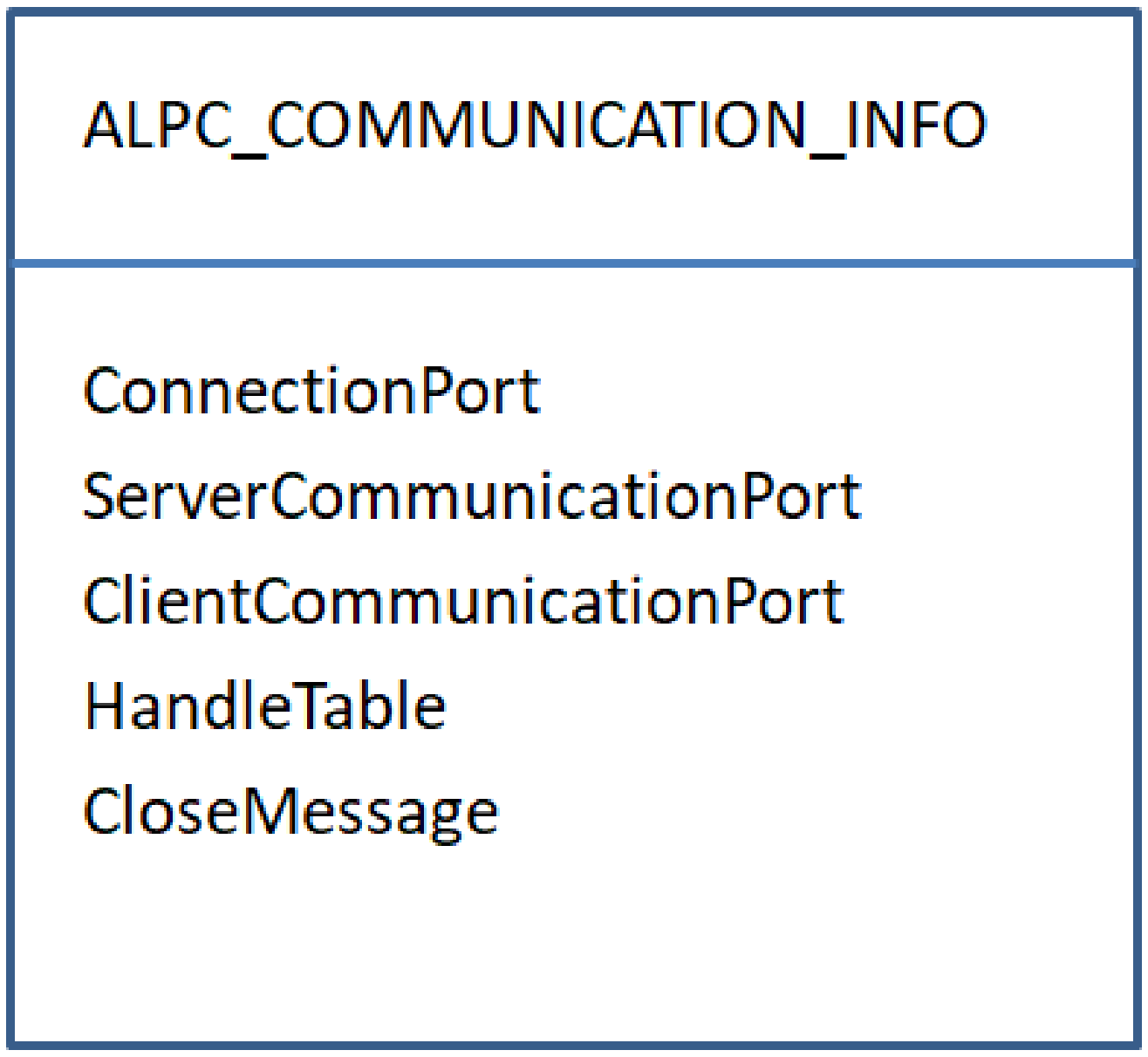

Figure 4. Fields of the ALPC_COMMUNICATION_INFO structure

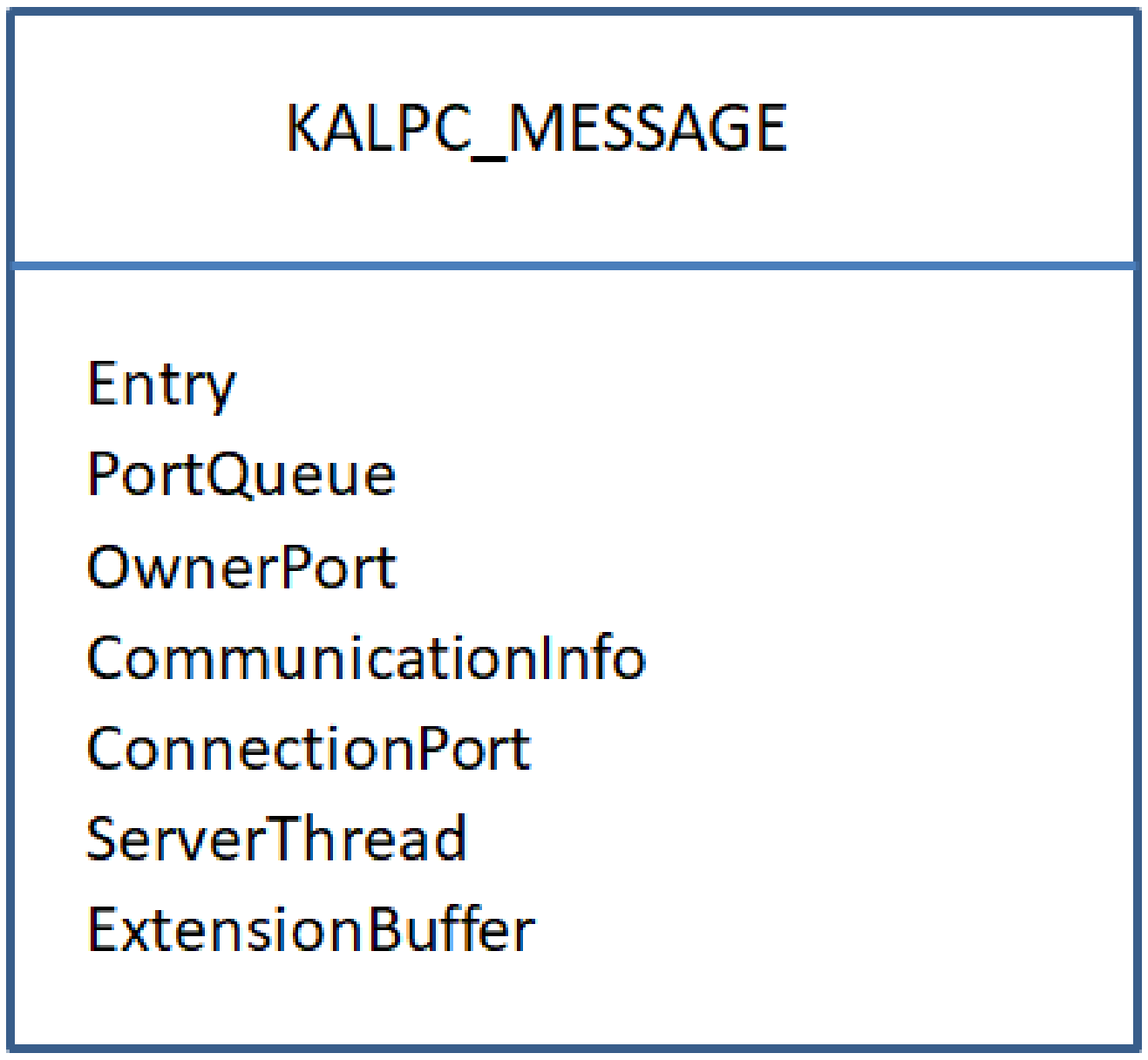

Figure 5. Fields of the KALPC_MESSAGE structure

### 2.4. Examples of programs for the enumeration of ALPC ports

This section describes programs that allow you to get information about the ALPC ports in the system.

**WinObj tool**

The objects of the ALPC port in the Windows system can be viewed using the WinObj tool from Sysinternals (Russinovich , 2022), which provides access to the namespace of the object manager.

Windows Internals book (Allievi, 2021) contains experiment of viewing subsystem ALPC port objects

**NtQuerySystemInformation function for enumeration of ALPC connections**

The function NTQuerySystemInformation is implemented on NTDLL (Microsoft, 2021). It basically retrieves specific information from the environment. This function can be used to retrieve numerous data from user mode, but no addresses of ALPC Connection and Communication Ports can be retrieved.

**Capabilities of the !alpc instruction**

The WinDBG debugger !alpc instruction provides a powerful tool for exploring ALPC ports from kernel mode perspective.

The instruction can be used with the following options:

!alpc /m <Message Address> dumps the message at the specified address;

!alpc /p <Port Address> dumps the port at the specified address;

!alpc /lpc <Port Address> dumps all connections for the specified port;

.

!alpc /lpp <Process Address> dumps all the connections for the specified process.

We run the Client-1 and Server-1 processes, interacting via ALPC, and used the /lpp option to collect information about ALPC client and server processes.

Firstly, the client process was explored. Client process has two client connections: the first with csrss.exe, the second – with the ALPC server, see Figure 6, *a)*. Figure 6, *b)* shows raw !alpc execution results for Client-1 process in WinDBG debugger. Let’s look closer the client connection to the ALPC server, indicated by the number 1. The first memory value in the connection string is the address of the ClientCommunicationPort structure, the second is the ConnectionPort address – the connection port, the third is the address of the ALPC server process itself.

Then Server-1 ALPC ports and connections were examined using the same instruction. A port that has been created by the server process is the Connection Port. Server process is connected to client process via the ServerCommunicationPort, see Figure 7, *a)*. Figure 7, *b)* shows raw !alpc execution results for the Server-1 process in WinDBG debugger. The name of the Connection Port is indicated in parentheses, it was set in the program code. The line below describes the ALPC connection with the client: the first address is the port address of the server itself (ServerCommunicationPort), then the client port address (ClientCommunicationPort) and the last value is the address of the client process.

The connection of the server to the Windows subsystem (csrss.exe) is indicated below.

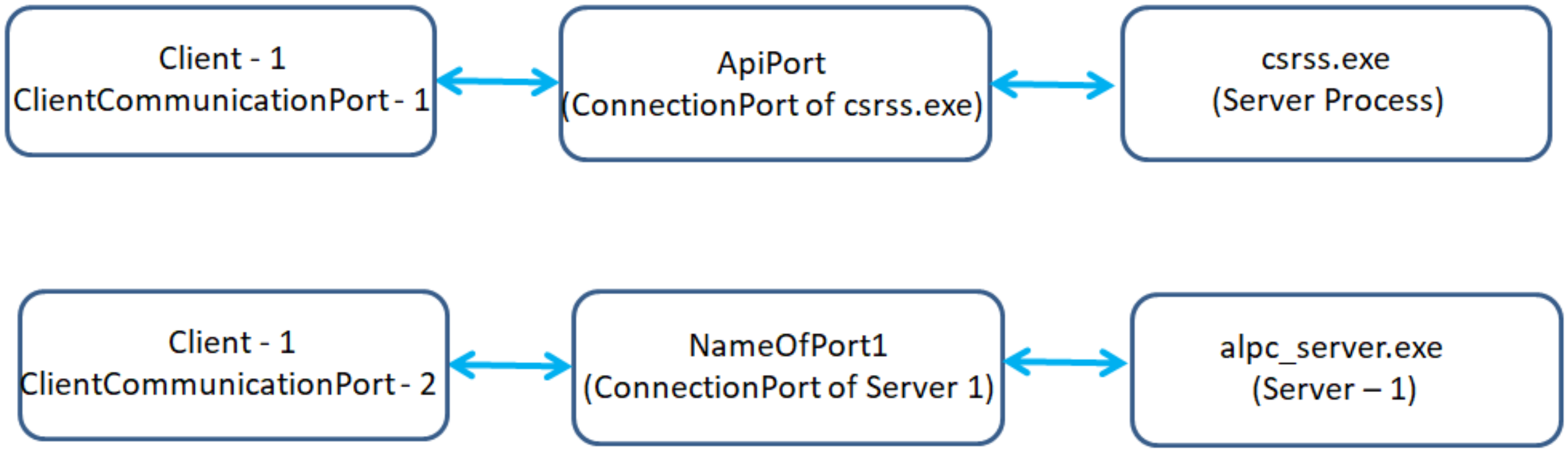

*a)*

```
kd> !alpc /lpp ffffa109`f4011080

Ports created by the process ffffa109f4011080:

        <none>

Ports the process ffffa109f4011080 is connected to:

        ffffa109f25f9a80 0 -> ffffa109f18c8d30 ('ApiPort') 0 ffffa109f18e41c0 ('csrss.exe')
        ffffa109f20f2d20 1 -> ffffa109f19da350 ('NameOfPort1') 1 ffffa109f31a8080 ('alpc_server.ex')
```

*b)*

Figure 6. Client Process Connections a) schematic results b) raw results

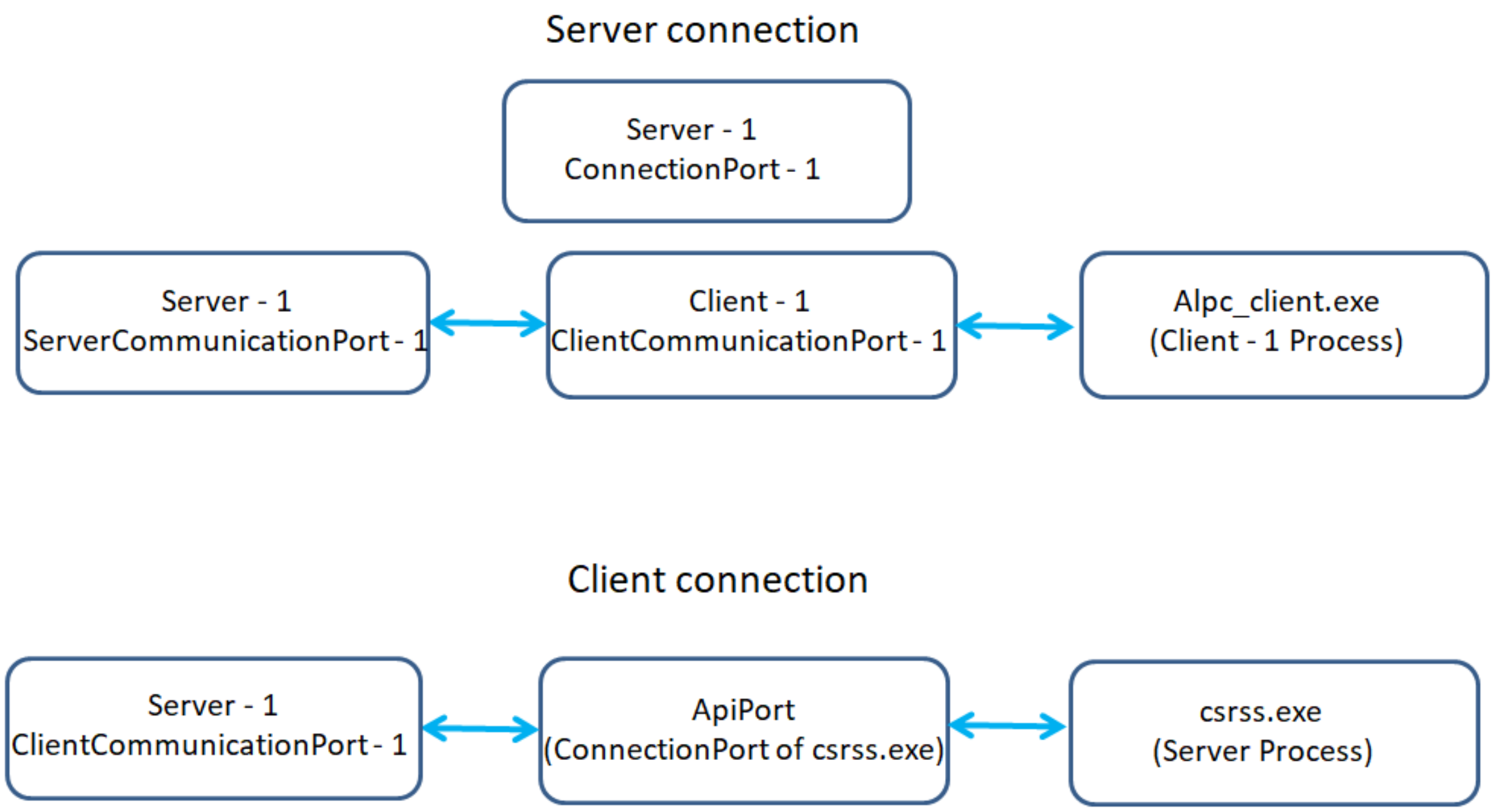

*a)*

```
kd> !alpc /lpp ffff8403a6b77080

Ports created by the process ffff8403a6b77080:

    ffff8403a64ccdb0('NameOfPort1') 1, 1 connections
        ffff8403a64d12d0 0 ->ffff8403a64d1530 1 ffff8403a64c8080('alpc_client.exe)

Ports the process ffff8403a6b77080 is connected to:

    ffff8403a5430ce0 0 -> ffff8403a40b4b60 ('ApiPort') 0 ffff8403a40b81c0 ('csrss.exe')
```

*b)*

Figure 7. Server Process Connections a) schematic results b) raw results

### 2.5. Possible vectors of attacks on the ALPC and their possible consequences

Based on the data obtained, it is possible to make assumptions about possible attack vectors on ALPC structures. ALPC interaction can be attacked via kernel drivers. Basic information about interacting ALPC structures is stored in the CommunicationInfo field of the ALPC port structure. Modification of the data in this structure can lead to a change in the objects of interaction or termination of it. An attack on ALPC interaction can be directed both at the client process and the client communication port belonging to it, and at the server process and the server communication port and connection port belonging to it.

## 3. RECENT RESEARCH PAPERS ON ALPC ATTACKS

At the LABScon 2022 Information security conference, researchers from the Binarly team Matrosov and Teodorescu drew attention to a new type of vulnerabilities that allows disabling the Windows management tool, Windows Management Instrumentation (WMI), without causing alarm (Matrosov, Teodorescu, 2022).

The researchers continued to reveal the topic they raised at the Black Hat 2022 conference (Teodorescu, Golchikov, Korkin, 2022a), where the Binarly Research Team demonstrated eight attacks on the WMI system and proved that the delivery of WMI events can be disabled, and there will be no security alerts.

### Interaction of WMI components via ALPC mechanism

WMI interacts with providers, clients, and the process services.exe through the ALPC mechanism.

The WMI service "Winmgmt" works as SVCHost process and acts as a server process in ALPC interaction, WMI clients act as ALPC clients and receive WMI events, the process services.exe communicates directly with the WMI service via the ALPC channel.

Process services.exe creates a server connection port named "\netsvcs" and receives a connection request from the WMI service.

After acceptance services.exe after a connection request, the WMI service receives the client's communication port descriptor, and services.exe receives a connection port descriptor, and an ALPC channel is established between them.

The WMI service creates a named connection port with the prefix "\RPC Control\OLE*" and receives connection requests from WMI clients.

Two possible attacks on ALPC interaction conducted from user mode were presented at the conference. The first attack blinded the WMI client by closing the handle of the client's ALPC communication port (Teodorescu, Golchikov, Korkin, 2022b). As a result of the attack, the connection was broken, and the client stopped receiving messages (Figure 8, a).

The second attack targeted the server side, closing the ALPC port descriptor on the server (Figure 8, b). After this attack, the server lost connection with all its clients (Teodorescu, Golchikov, Korkin, 2022c). All attempts to connect to it ended with error 0x800706BF (Remote Procedure Call error).

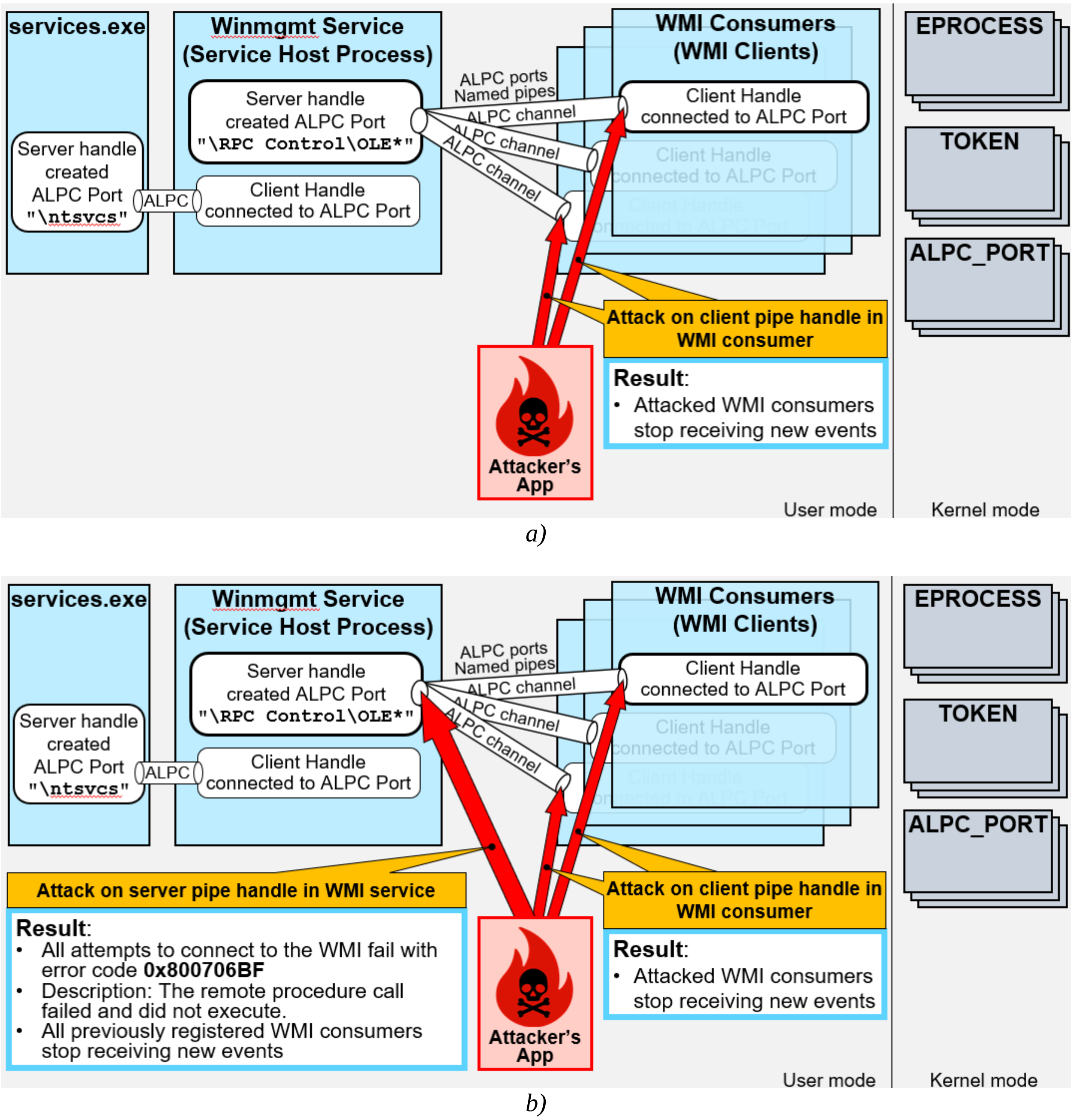

Figure 8. Scheme of attack on the ALPC port descriptor of the *a) client process b) server process*

**Kernel-mode ALPC attack**

At the Ekoparty 2022 conference (Teodorescu, Korkin, 2022) Binarly Research Team researchers continued their research of WMI vulnerabilities and presented various vectors of possible attacks on the Windows management tool — Windows Management Instrumentation (WMI), among which several attacks on ALPC interaction are being considered and conducted.

An attack on the ALPC port of the application, changing the address of the TLS global context to -1, was demonstrated. As a result, ALPC connection was disrupted.

Kernel mode attack on ALPC was also showed at the conference, see Figure 9. The attacker installs the kernel driver and uses this driver to reset the structure of the ALPC port in the memory of the client and server processes.

As a result, the ALPC connection is broken, and the client and server cannot restore it and create a new connection.

Thus, at the Ekoparty and LABScon conferences in 2022, it was shown that WMI is vulnerable to attacks from both user mode and kernel mode. Attacks on the ALPC connection that have not been investigated before pose a great danger.

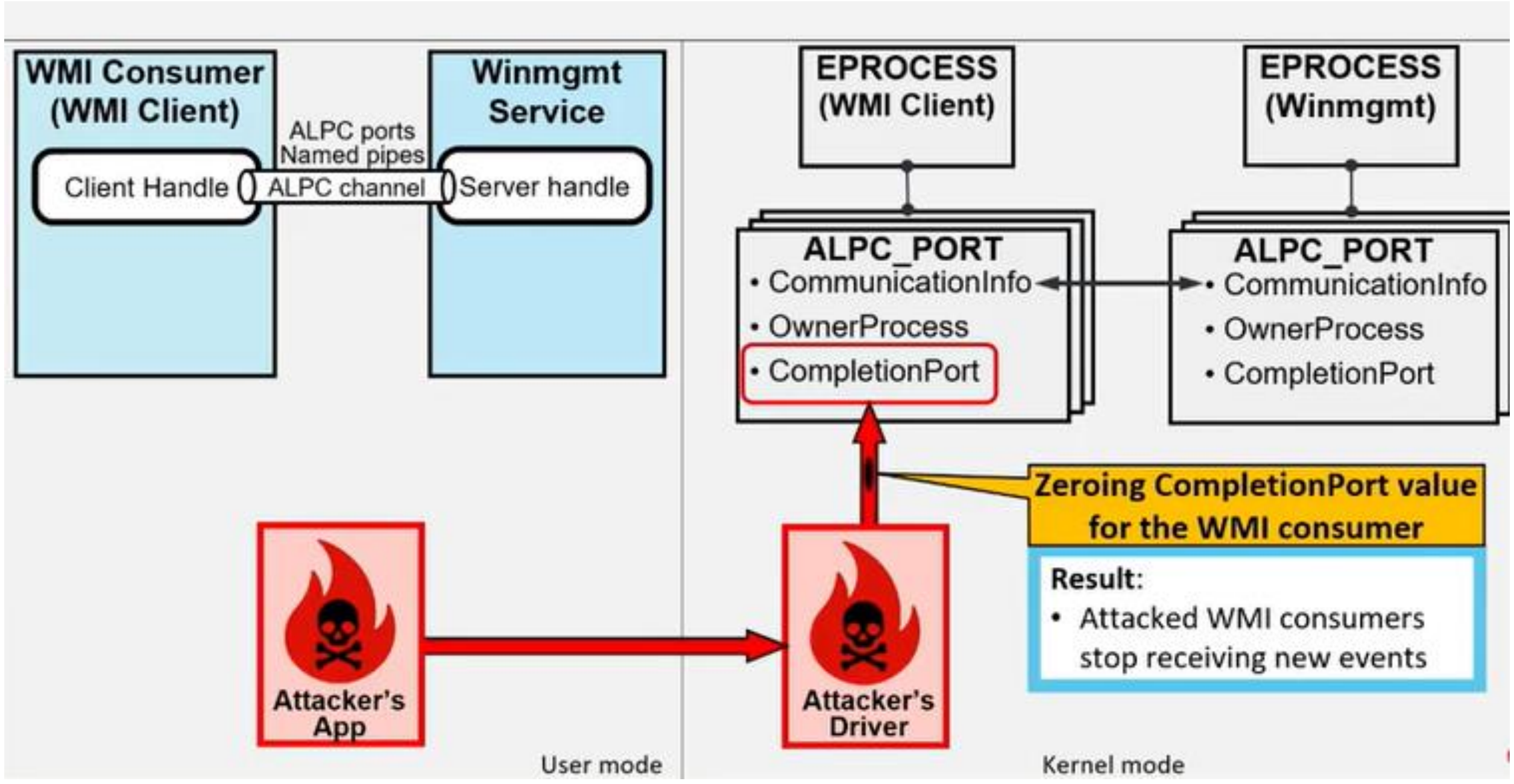

Figure 9. Scheme of kernel mode attack on the ALPC port structure

## 4. NEW ATTACKS ON ALPC USING THE KERNEL DRIVER

This section describes the attacks on ALPC, conducted using the kernel driver.

### 4.1. Test bench

To conduct the research, we created a bench consisting of applications client C1 and server S1 interacting via ALPC and client C2 and server S2 interacting via ALPC, see Figure 10.

The ProcessHacker APK header file (ntlpcapi.h) was used as the header file.

The client sends a message to the server, and the server, after receiving the message, outputs this message to the console.

We run processes C1, S1, C2 and S2 on the virtual machine Windows 11 and made sure that server S1 received messages from client C1, and server S2 received messages from client C2.

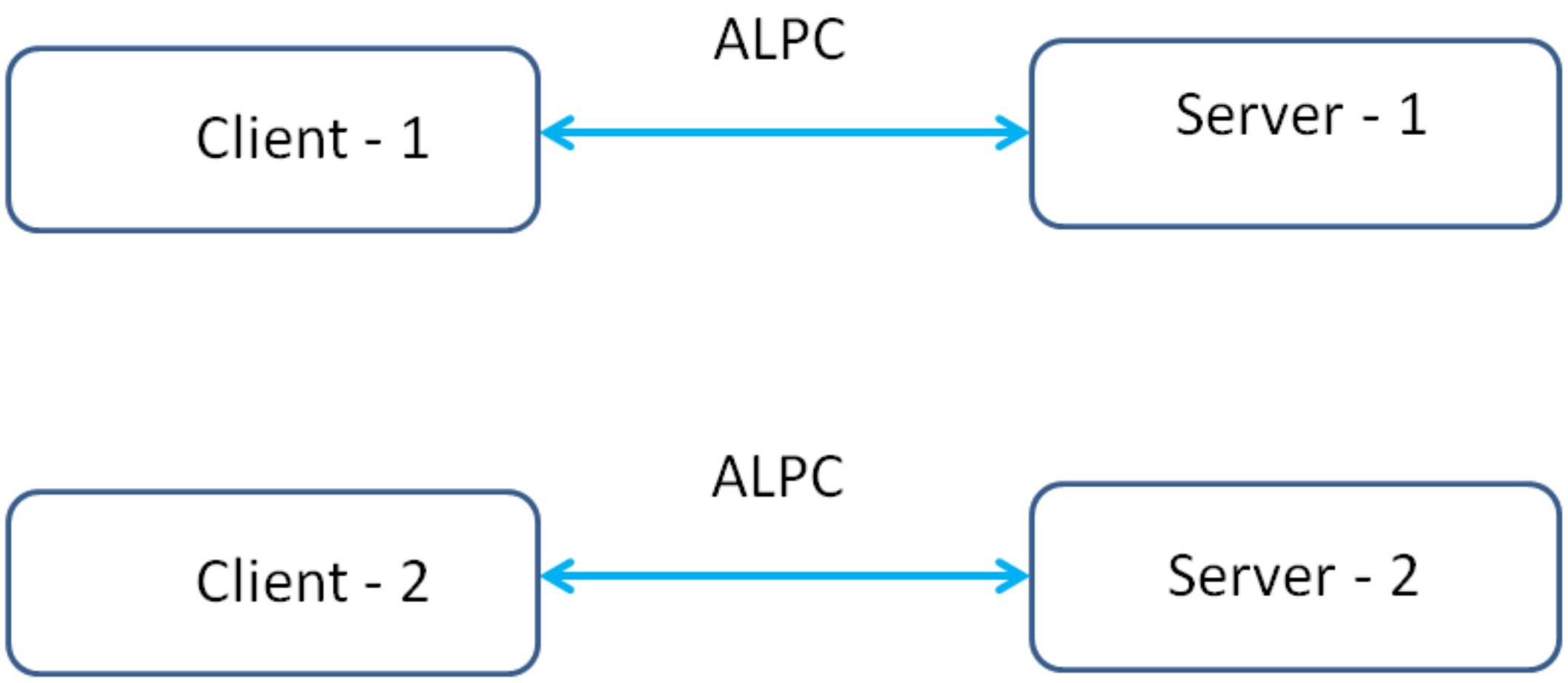

Figure 10 — ALPC connections of the bench objects

The virtual machine was connected to WinDBG. We executed !dml_proc instruction to get process addresses of the running processes and used described in 2.3 !alpc instruction to get the addresses of ALPC port structures.

Dynamically modifying the values of the ALPC ports fields in the Watch WinDBG tab, we carried out the following attacks.

### 4.2. Attack №1 – spoofing and blinding

Attack №1 is a spoofing attack for Server S1 and blinding attack for S2. This attack targets client side. We change the value of Connection Port in the CommunicationInfo structure of the Client Communication Port of client C2 to the address of the Connection Port structure of ConnectionPort C1.

The attack scheme is shown in Figure 11.

After starting the machine, we see that server S1 now receives messages from client C2, which is not officially connected to it. Server S2 is blinded and does not receive any messages.

Client C1 can also send messages to S1.

Thus, the spoofing and blinding attack №1 on the ALPC connection was successfully carried out. Client C2 began to send messages to the message queue of S1 because of the replacement of the Connection Port address in the Client Communication Port structure. C1 also has the address of this queue and continues to send messages to it.

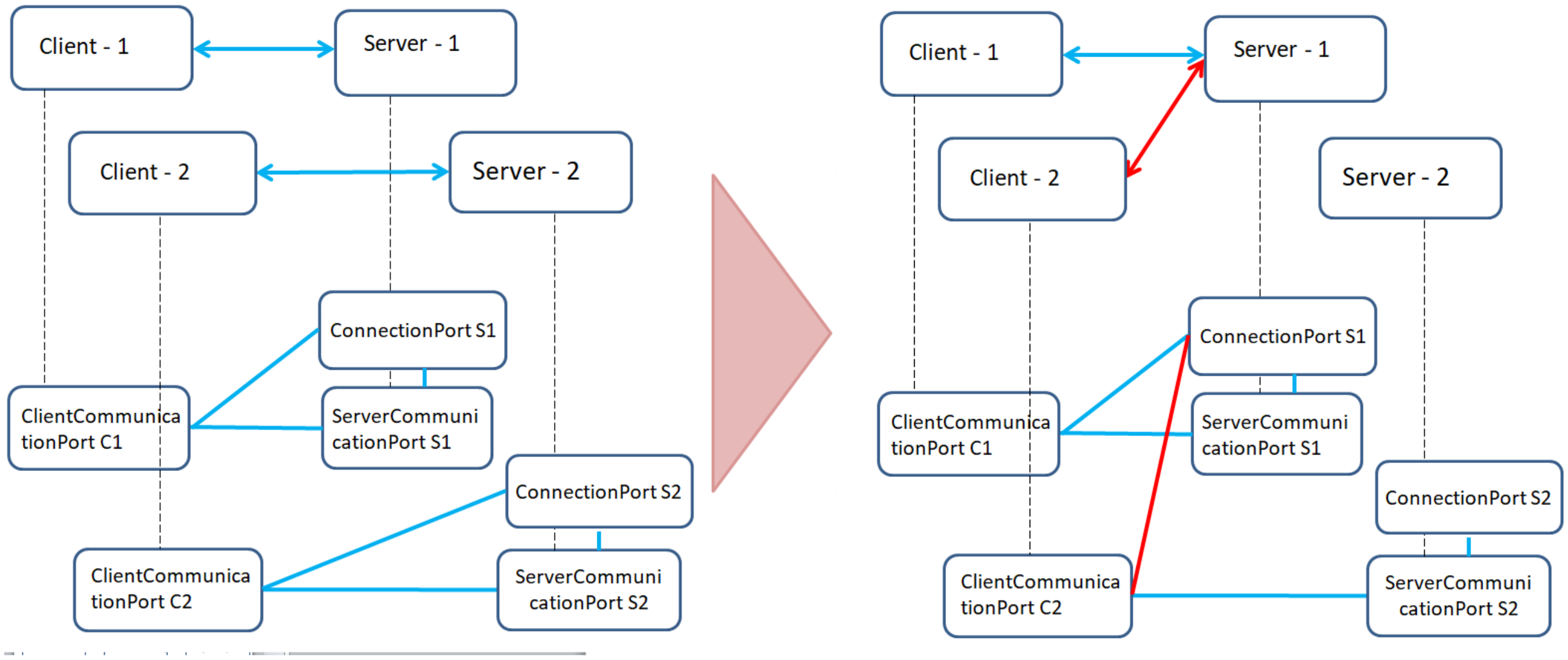

Figure 11. Attack №1 scheme

### 4.3. Attack №2 – spoofing attack

Attack №2 also targets the client's side. The attack is performed in three stages.

Firstly, attack №1 is conducted.

Secondly, blinded server S2 is terminated. After the attack №1 the message about the wrong ALPC connection can be found by the !alpc instruction in the records of the blinded server. Therefore, it may be important for an attacker to shut it down after carrying out attack № 1 in order to hide his malicious activity.

Lastly, the ServerCommunicationPort value in the same CommunicationInfo structure of the C2 Client Communication Port is replaced with the S1 Server Communication Port address.

Third stage does not lead to result, when it is performed alone. Still, it is essential part of the attack №2. After the attack №1 to the C1-S1 connection client C2 connected to server S1, but after server S2 was completed, client C2, who had already connected to server S1, stopped sending messages to it, receiving error 0xC0000037.

In the debugger we can see that the value of Server Communication Port in the CommunicationInfo field of the Client Communication Port C2 has been reset to zero (Fig. 13).

After we performed the replacement of the zeroed ServerCommunicationPort field the connection C2 to S1 was restored.

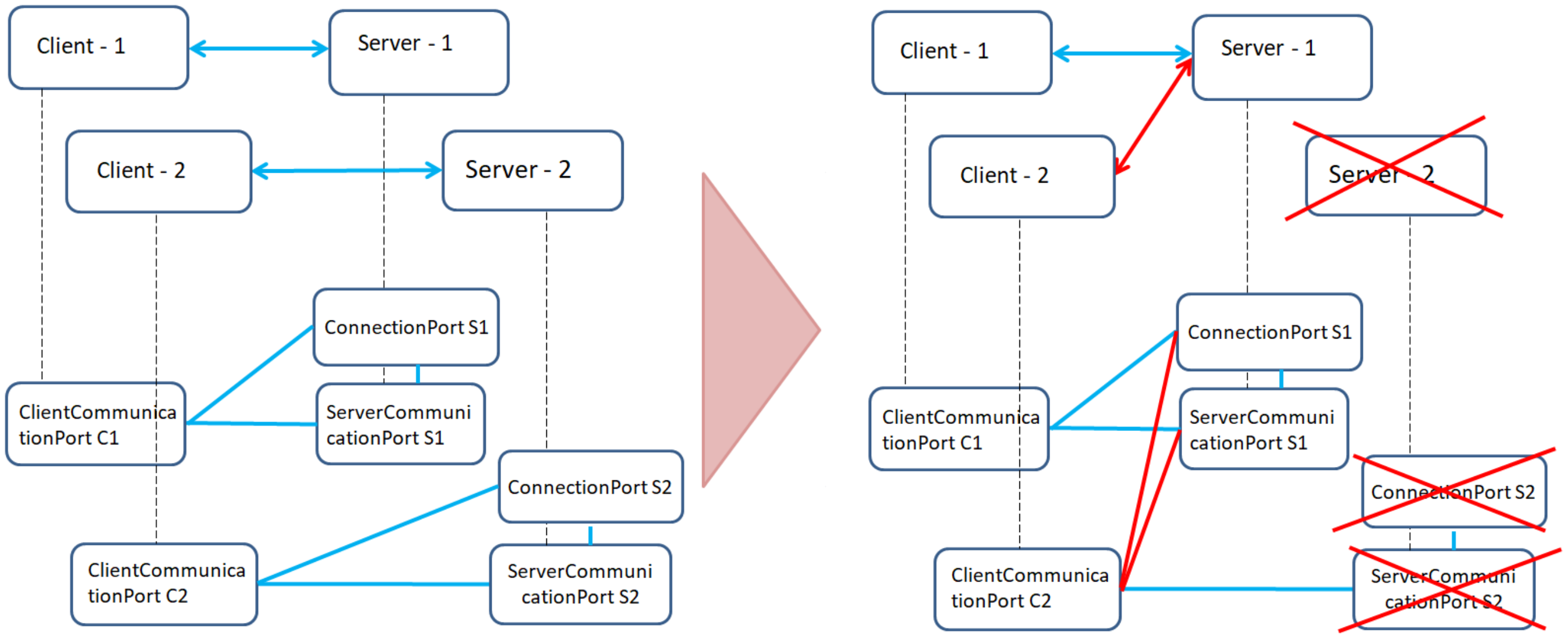

Figure 12. Attack №2 scheme

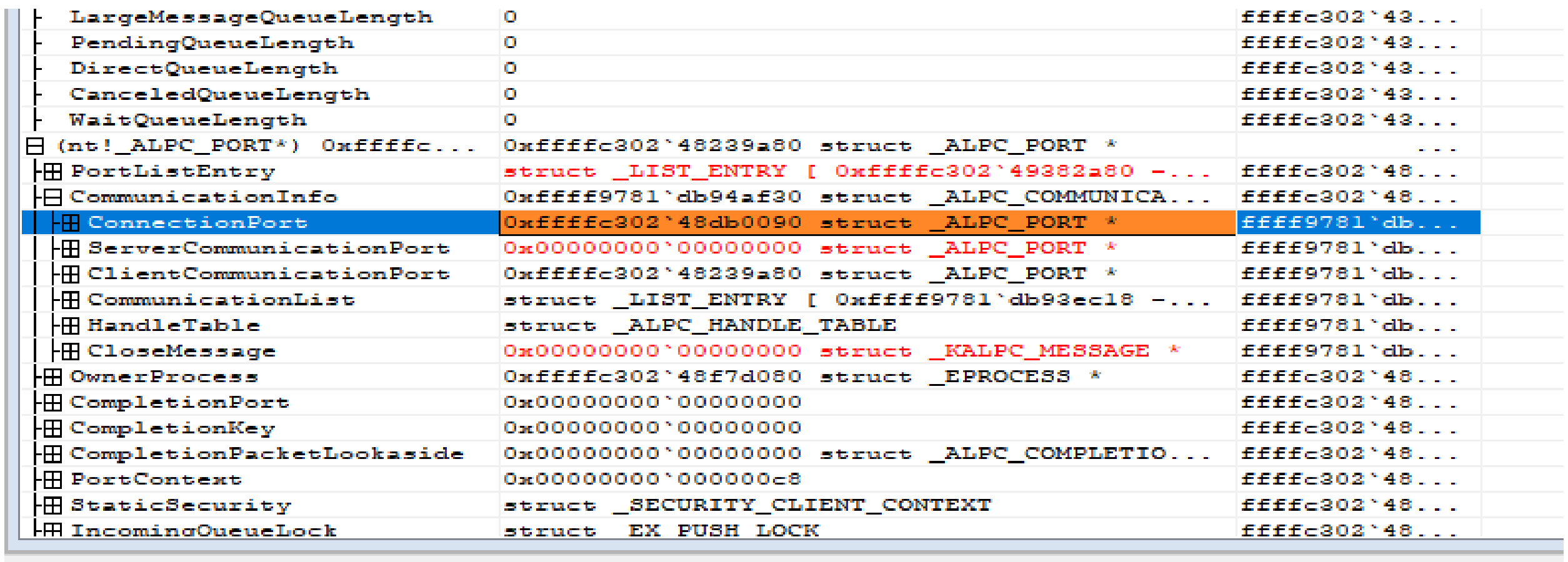

| | | |
|---|---|---|
| LargeMessageQueueLength | 0 | ffffc302`43... |
| PendingQueueLength | 0 | ffffc302`43... |
| DirectQueueLength | 0 | ffffc302`43... |
| CanceledQueueLength | 0 | ffffc302`43... |
| WaitQueueLength | 0 | ffffc302`43... |
| (nt!_ALPC_PORT*) 0xffffc... | 0xffffc302`48239a80 struct _ALPC_PORT * | ... |
| PortListEntry | struct _LIST_ENTRY [ 0xffffc302`49382a80 -... | ffffc302`48... |
| CommunicationInfo | 0xffff9781`db94af30 struct _ALPC_COMMUNICA... | ffffc302`48... |
| ConnectionPort | 0xffffc302`48db0090 struct _ALPC_PORT * | ffff9781`db... |
| ServerCommunicationPort | 0x00000000`00000000 struct _ALPC_PORT * | ffff9781`db... |
| ClientCommunicationPort | 0xffffc302`48239a80 struct _ALPC_PORT * | ffff9781`db... |
| CommunicationList | struct _LIST_ENTRY [ 0xffff9781`db93ec18 -... | ffff9781`db... |
| HandleTable | struct _ALPC_HANDLE_TABLE | ffff9781`db... |
| CloseMessage | 0x00000000`00000000 struct _KALPC_MESSAGE * | ffff9781`db... |
| OwnerProcess | 0xffffc302`48f7d080 struct _EPROCESS * | ffffc302`48... |
| CompletionPort | 0x00000000`00000000 | ffffc302`48... |
| CompletionKey | 0x00000000`00000000 | ffffc302`48... |
| CompletionPacketLookaside | 0x00000000`00000000 struct _ALPC_COMPLETIO... | ffffc302`48... |
| PortContext | 0x00000000`000000c8 | ffffc302`48... |
| StaticSecurity | struct _SECURITY_CLIENT_CONTEXT | ffffc302`48... |
| IncomingQueueLock | struct _EX_PUSH_LOCK | ffffc302`48... |

Figure 13. The fields of the Client Communication Port structure of the client C2 after the completion of the server S2

### 4.4. Attack №3 – spoofing and blinding

Attack №3 targets the server side. The attack imposes illegitimate connection to Server S1 and blinds S2.

The ConnectionPort value in the CommunicationInfo field of the Server Communication Port of server S1 was changed to the Connection Port address of the S2 Server Communication Port (Figure 14).

Thus, server S1, while continuing to store the Connection Port descriptor S1, began to take messages from the message queue of Connection Port S2, where client C2 sends messages. At the same time, messages sent by client C1 come to the Connection Port S1, the handle of which is stored in the server process, and messages C1 also get into the message stream and come to server S1.

As a result, Server S1 receives all messages from clients C1 and C2.

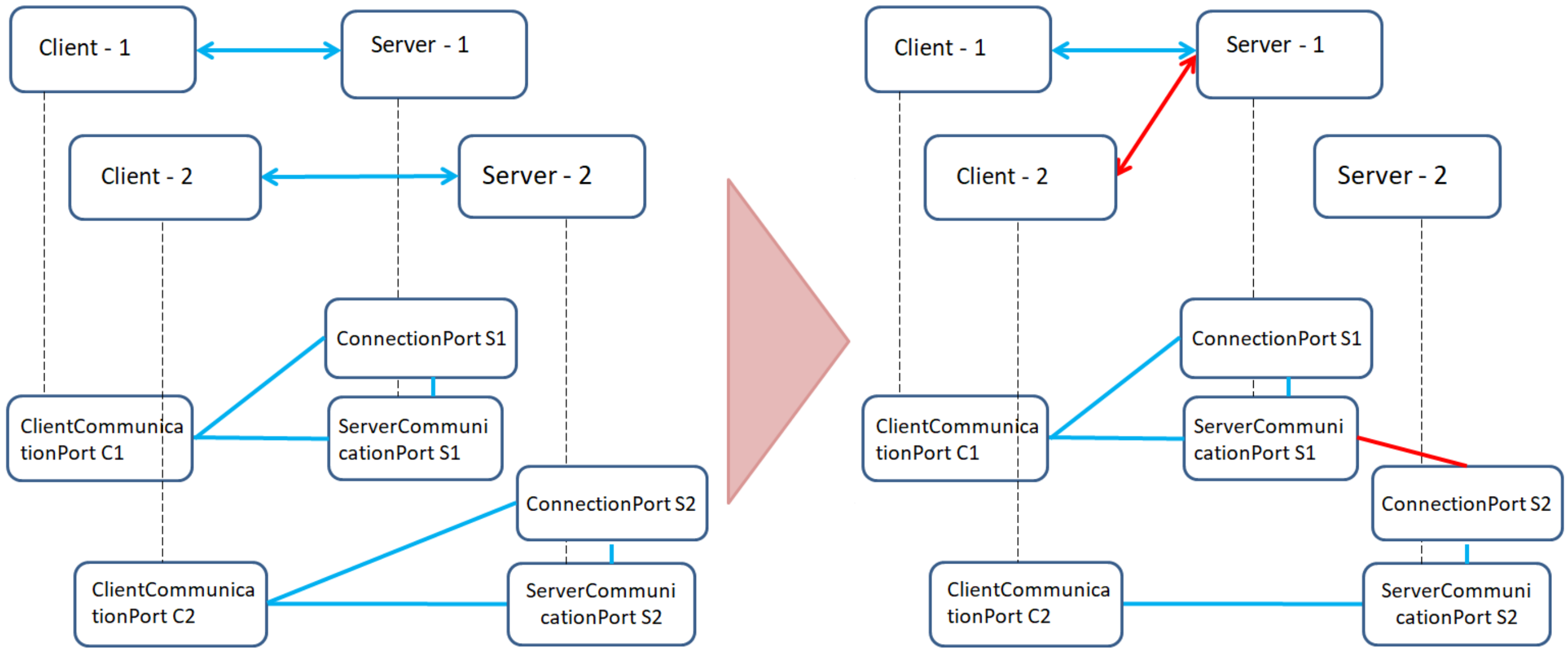

Figure 14. Attack №3 scheme

### 4.5. Attacks results

Three attacks on the ALPC connection were demonstrated in this section. Their results are presented in the Table 1.

Table 1 Attacks on ALPC and their results

| Attack № | Involved apps | Changeable fields | Result |
|---|---|---|---|
| 1 | Client C2 | ClientCommunicationPort->CommunicationInfo->ConnectionPort | Spoofing attack on S1, illegitimate connection C2-S1, server S2 blinded |
| 2 | Client C2 | ClientCommunicationPort->CommunicationInfo->ConnectionPort, ClientCommunicationPort->CommunicationInfo->ServerCommunicationPort | Spoofing attack on S1, illegitimate connection C2-S1, no messages about wrong connection can be found using !alpc instruction |
| 3 | Server S1 | ServerCommunicationPort->CommunicationInfo->ConnectionPort | Spoofing attack on S1, illegitimate connection C2-S1, server S2 blinded |

## 5. ALPCHECKER DETECTS SPOOFING AND BLINDING

### 5.1. ALPC attack identifiers

When attack №1 or attack №3 is accomplished, the line "ConnectionPort for _ALPC_COMMUNICATION_INFO <CommunicationInfo> points to wrong port <PortAddress>" will be among the information output by the !alpc /lpp <blinded_server_address> instruction, see Figure 15. This message can be used as ALPC attack identifier.

Attack №2 terminates the blinded process, so there is no such message in the ALPC logs.

In the correctly functioning system each client process contains information about its ALPC connections and each server process also contains information about its ALPC connections to clients. Therefore, the same ALPC information about the Connection Port, Client Communication Port and Server Communication Port is stored by both the client and server.

When an attacker modifies ALPC data, mismatched ALPC information appears. Thus, if there is client connection for which there is no server connection with the same ALPC ports, it can be attack identifier. Such connection is defined as 'suspicious connection'.

```
Ports created by the process ffffad83e845d2c0:

    ffffad83e788c090('NameOfPort2') 1, 1 connections
  ConnectionPort for _ALPC_COMMUNICATION_INFO ffffcb8c7c8d20d0 points to wrong port ffffad83e71b8a00
         ffffad83e84ccd90        0       ->ffffad83e84aad90        1 ffffad83e8559080('alpc_client2.e')
```

Figure 15. ALPC information of the blinded process

### 5.2. ALPChecker algorithm

We developed ALPChecker – tool that detects attack on the ALPC. ALPChecker checks the living system for the presence of attack identifiers and warns the user of danger.

First, the script collects information about all running processes using the debugger command "!dml_proc". Then the instruction "!alpc /lpp <ProcessAddress>" is executed for each process, the received information is saved in a log file. Information about server and client connections is selected from this file, sorted and stored in memory. If the message about connection to a wrong port is found, script prints alert message. Next, for each client connection, the existence of a server process with identical data about processes and ALPC ports is checked. If a mismatch is found, it may be a sign of an attack, and a message is displayed with information about this connection.

The flowchart of the algorithm is presented in the figure 16.

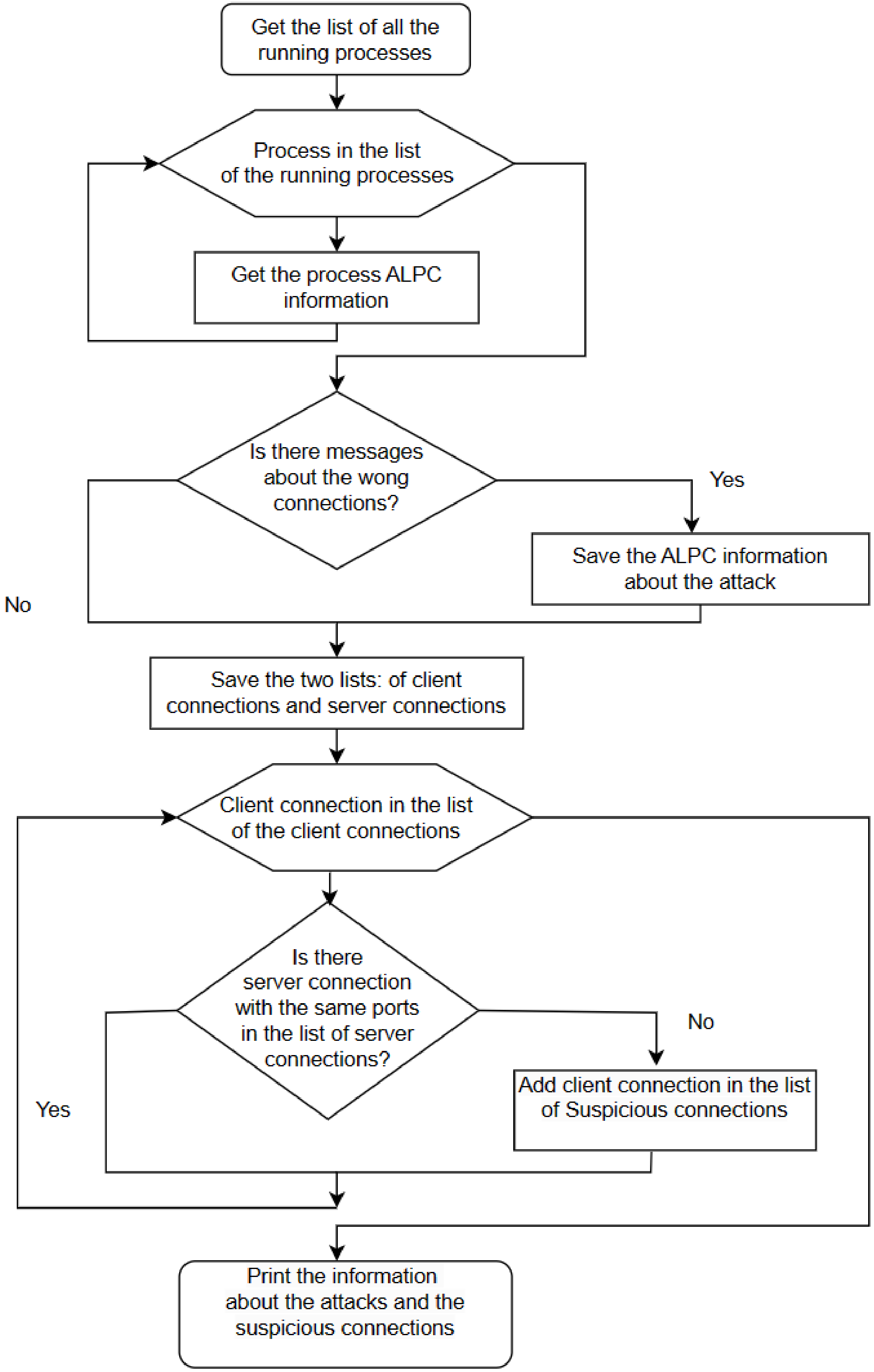

Figure 16. ALPChecker flowchart

### 5.3. ALPChecker architecture

ALPChecker is written in Python using the os, sys, and subprocess libraries. This tool uses livekd in order to work in user mode, but collect and analyse the information, available in kernel mode. Livekd is a user mode tool for Windows that uses its driver livekdd.sys to get kernel-mode system information. Livekd also allows you to use the KD and WinDBG debuggers from the Debugging Tools for Windows package locally from a running operating system (Russinovich, Johnson, 2020). All debugging commands are available in the livekd to get information about the internal processes of the system.

ALPChecker is presented in the Figure 17.

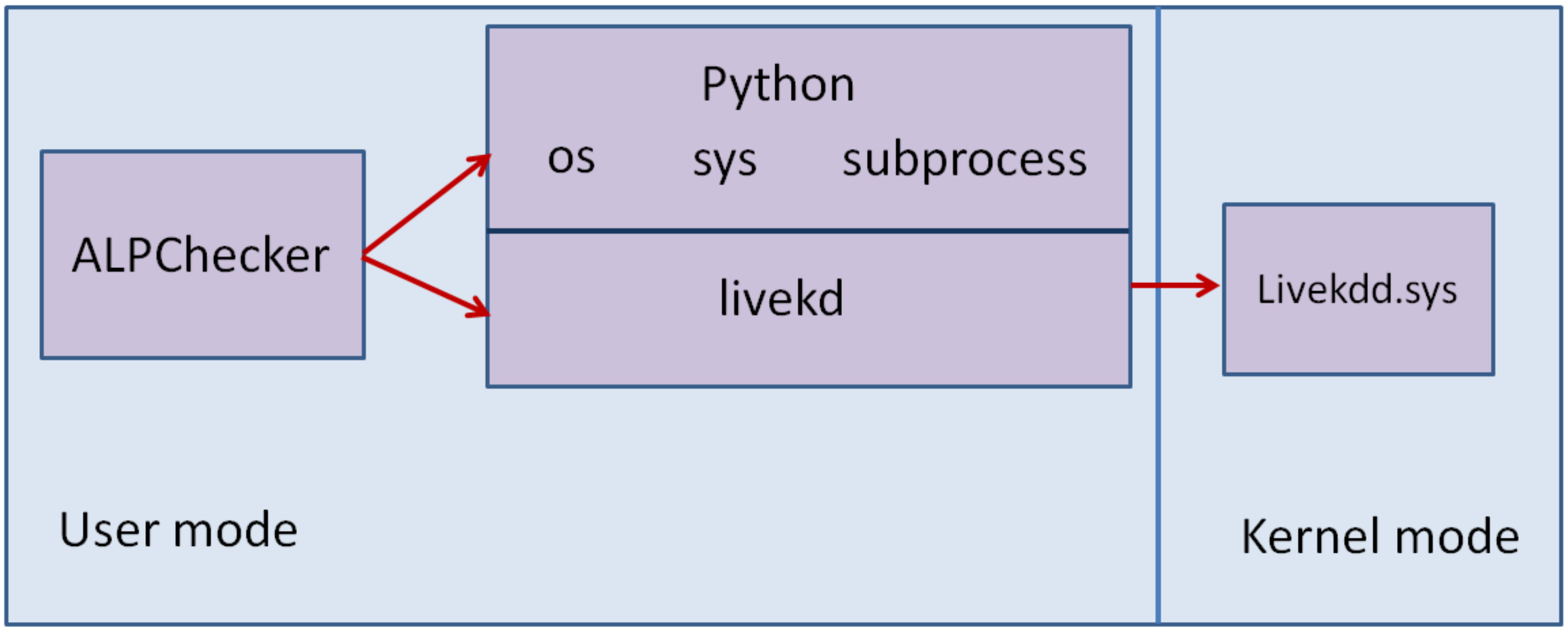

Figure 17. ALPChecker architecture

### 5.4. Execution of ALPChecker

When the program had been launched, it was discovered that in a correctly functioning system may be several (most often from two to five) client connections, about which the server process has no information.

The next time we run the program there were no such processes with such ports, therefore these were just system processes captured at the time of completion or unsuccessful attempts of connection.

Thus, we can not be absolutely sure that all 'suspicious connections' indicate an attack. User can run the program again to verify the status of the connection.

ALPChecker outputs a large amount of information for user to make a conclusion about the connection: name, pid and addresses of client and server processes, user and path of the client processes, name, address and the list of connections of ConnectionPort, address of the Client Communication Port.

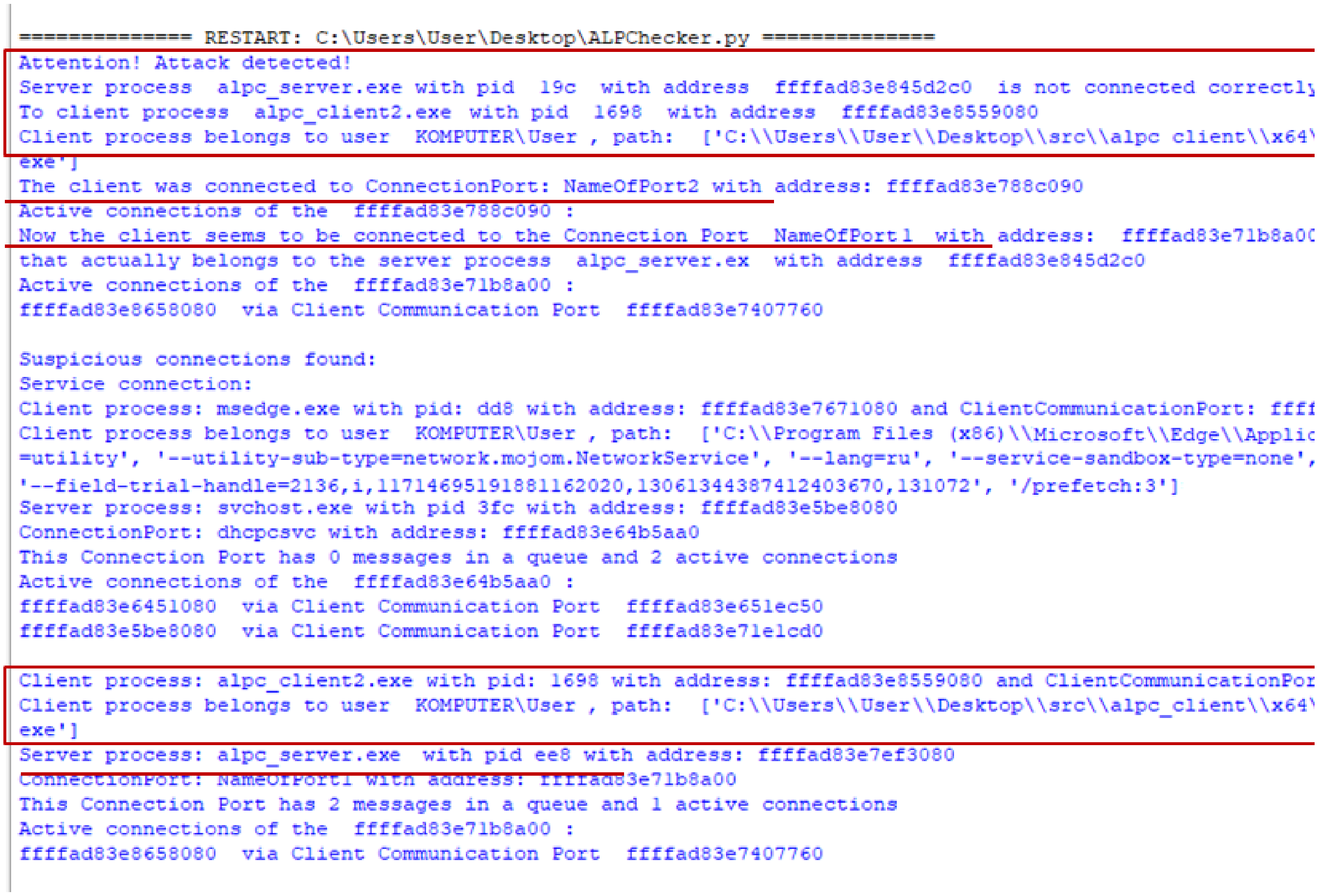

```
============== RESTART: C:\Users\User\Desktop\ALPChecker.py ==============
Attention! Attack detected!
Server process  alpc_server.exe with pid  19c  with address  ffffad83e845d2c0  is not connected correctly
To client process  alpc_client2.exe with pid  1698  with address  ffffad83e8559080
Client process belongs to user  KOMPUTER\User , path:  ['C:\\Users\\User\\Desktop\\src\\alpc client\\x64\
exe']
The client was connected to ConnectionPort: NameOfPort2 with address: ffffad83e788c090
Active connections of the  ffffad83e788c090 :
Now the client seems to be connected to the Connection Port  NameOfPort1  with address:  ffffad83e71b8a00
that actually belongs to the server process  alpc_server.ex  with address  ffffad83e845d2c0
Active connections of the  ffffad83e71b8a00 :
ffffad83e8658080  via Client Communication Port  ffffad83e7407760

Suspicious connections found:
Service connection:
Client process: msedge.exe with pid: dd8 with address: ffffad83e7671080 and ClientCommunicationPort: ffff
Client process belongs to user  KOMPUTER\User , path:  ['C:\\Program Files (x86)\\Microsoft\\Edge\\Applic
=utility', '--utility-sub-type=network.mojom.NetworkService', '--lang=ru', '--service-sandbox-type=none',
'--field-trial-handle=2136,i,11714695191881162020,13061344387412403670,131072', '/prefetch:3']
Server process: svchost.exe with pid 3fc with address: ffffad83e5be8080
ConnectionPort: dhcpcsvc with address: ffffad83e64b5aa0
This Connection Port has 0 messages in a queue and 2 active connections
Active connections of the  ffffad83e64b5aa0 :
ffffad83e6451080  via Client Communication Port  ffffad83e651ec50
ffffad83e5be8080  via Client Communication Port  ffffad83e71e1cd0

Client process: alpc_client2.exe with pid: 1698 with address: ffffad83e8559080 and ClientCommunicationPor
Client process belongs to user  KOMPUTER\User , path:  ['C:\\Users\\User\\Desktop\\src\\alpc_client\\x64\
exe']
Server process: alpc_server.exe  with pid ee8 with address: ffffad83e7ef3080
ConnectionPort: NameOfPort1 with address: ffffad83e71b8a00
This Connection Port has 2 messages in a queue and 1 active connections
Active connections of the  ffffad83e71b8a00 :
ffffad83e8658080  via Client Communication Port  ffffad83e7407760
```

Figure 18. ALPChecker detects the attack

After attack №1 and attack №3 ALPChecker detects the attack by the message about incorrect connection in the ALPC logs and prints the alert message, see Figure 18. The suspicious connection C2-S1 can be found among the suspicious connections below.

As noted in 4.3, attack №2 includes terminating the blinded process. Therefore, there is no message about wrong port in the log file. But ALPChecker still finds the suspicious connection C2-S1 and outputs its information.

## 6. CONCLUSION

To sum up we would like to highlight the following:

1. Today ALPC mechanism is widely used in Windows OS. Even a simple program has at least one ALPC connection. Windows does not provide security tools to defend ALPC from kernel mode attacks.
2. Three new attacks on ALPC were presented. Attack №1 was spoofing and blinding attack, changing ConnectionPort value in the CommunicationInfo field of the Client Communication Port structure of the attacker client process. As a result of the attack the client got unauthorized access to the ALPC server. Sever S2 was blinded. Attack №2 was spoofing attack, changing ConnectionPort and ServerCommunicationPort values in the CommunicationInfo field of the Client Communication Port structure of the attacker client process and terminating blinded process. Attack №3 was spoofing and blinding attack, changing the ConnectionPort value in the CommunicationInfo field of the Server

Communication Port structure of the attacker server process. As a result of the attack, server process got all the messages that were sent to the legitimate server.
3. Two factors can be used to detect the attack: ALPC record that a field in ALPC_COMMUNICATION_INFO structure points to wrong port and differences between client processes and server processes information.
4. ALPChecker collects and analyses ALPC information and detects ALPC spoofing and blinding attacks.

## 7. AVAILABILITY OF DATA AND MATERIALS

Not applicable.

## 8. FUNDING

No financial support was made.

## 9. ACKNOWLEDGEMENTS

Not applicable.

## 10. REFERENCES

[1] Abrams, L. (2022). Malware now using NVIDIA's stolen code signing certificates. Retrieved from https://www.bleepingcomputer.com/news/security/malware-now-using-nvidias-stolen-code-signing-certificates/

[2] Allievi, A., Ionescu, A., Russinovich, M., Solomon, D. (2021). *Windows Internals* (7th ed.). Parts 1 and 2. Redmond, Washington: Microsoft Press.

[3] Arghire, I. (11 01 2023). Cybercrime Group Exploiting Old Windows Driver Vulnerability to Bypass Security Products. SecurityWeek. Retrieved from https://www.securityweek.com/cybercrime-group-exploiting-old-windows-driver-vulnerability-bypass-security-products

[4] Baines, J. (2021). Driver-Based Attacks: Past and Present. Retrieved from https://www.rapid7.com/blog/post/2021/12/13/driver-based-attacks-past-and-present/

[5] Barysevich, A. (2018). The Use of Counterfeit Code Signing Certificates Is on the Rise. Recorded Future. Retrieved from https://www.recordedfuture.com/code-signing-certificates/

[6] Chlumecký, M. (2021). DirtyMoe: Introduction and General Overview of Modularized Malware. Retrieved from https://decoded.avast.io/martinchlumecky/dirtymoe-1/

[7] ESET. (2018). LoJax: First UEFI rootkit found in the wild, courtesy of the Sednit group. Retrieved from https://www.welivesecurity.com/2018/09/27/lojax-first-uefi-rootkit-found-wild-courtesy-sednit-group/

[8] Freedesktop. (2022). D-Bus. Retrieved from https://www.freedesktop.org/wiki/Software/dbus

[9] Garnier, T. (2008). Windows privilege escalation through LPC and ALPC Interfaces. SkyRecon. Retrieved from https://recon.cx/2008/a/thomas_garnier/LPC-ALPC-paper.pdf

[10] Gupta, R., Dresel, L., Spahn, N., Vigna, G., Kruegel, K., Kim, T. (2022). POPKORN: Popping Windows Kernel Drivers At Scale. Retrieved from https://dl.acm.org/doi/pdf/10.1145/3564625.3564631

[11] Hfiref0x. (2019a). Driver Loader for bypassing Windows x64 Driver Signature Enforcement. Github. Retrieved from https://github.com/hfiref0x/TDL

[12] Hfiref0x. (2019b). Universal PatchGuard and Driver Signature Enforcement Disable. Github. Retrieved from https://github.com/hfiref0x/UPGDSED

[13] Hfiref0x. (2022).Kernel Driver utility. Github. Retrieved from https://github.com/hfiref0x/KDU

[14] Iacob I., Ionita, I. M. (2022). The anatomy of Wiper Malware, Part 2: Third-Party Drivers. CrowdStrike. Retrieved from https://www.crowdstrike.com/blog/the-anatomy-of-wiper-malware-part-2

[15] Ionescu, A. (2014). Ionescu, A. All About The Rpc, Lrpc, Alpc, And Lpc In Your Pc. Singapore. SyScan’14. Retrieved from

https://www.youtube.com/watch&v=UNpL5csYC1E
[16] Freedesktop. (2022). D-Bus. Retrieved from https://www.freedesktop.org/wiki/Software/dbus
[17] Klein, A., Kotler, I. (2019). Windows Process Injection in 2019. Blackhat 2019. Retrievd from https://i.blackhat.com/USA-19/Thursday/us-19-Kotler-Process-Injection-Techniques-Gotta-Catch-Them-All-wp.pd
[18] Korkin, I. (2021, May 24-27). Protected Process Light is not Protected: MemoryRanger Fills the Gap Again. Paper presented at the Systematic Approaches to Digital Forensic Engineering (SADFE) International Workshop in conjunction with the 42nd IEEE Symposium on Security and Privacy. in Proceedings of 2021 IEEE Symposium on Security and Privacy Workshops, San Francisco, CA, USA, May 24-27, 2021, pp.298-308, Retrieved from https://conferences.computer.org/sp/pdfs/spw/2021/893400a298.pdf doi.org/10.1109/SPW53761.2021.00050
[19] Lechtic, M. (2021). GhostEmperor: From ProxyLogon to kernel mode. Kaspersky Lab. Retrieved from https://www.haktechs.com/ghostemperor-from-proxylogon-to-kernel-mode
[20] Lechtik, M., Berdnikov, V., Legezo, D., Borisov, I. (2022). MoonBounce: the dark side of UEFI firmware. Kaspersky Lab. Retrieved from https://securelist.com/moonbounce-the-dark-side-of-uefi-firmware/105468/
[21] Magdy, S., Zohdy, M. (19 12 2022a). A Closer Look at Windows Kernel Threats. TrenMicro. Retrieved from https://www.trendmicro.com/en_us/research/22/l/a-closer-look-at-windows-kernel-threats.html
[22] Magdy, S., Zohdy, M. (2022b). An In-Depth Look at Windows Kernel Threats. TrendMicro. Retrieved from https://documents.trendmicro.com/assets/white_papers/wp-an-in-depth-look-at-windows-kernel-threats.pdf
[23] Magdy, S., Zohdy, M. (05 01 2023). The evolution of Windows kernel threats. TrendMicro. Retrieved from https://www.trendmicro.com/vinfo/us/security/news/cybercrime-and-digital-threats/the-evolution-of-windows-kernel-threats
[24] Matrosov, A. Teodorescu, C. (2022). New Attacks To Disable And Bypass Windows Management Instrumentation. LABScon 2022. Retrieved from https://binarly.io/posts/New_Attacks_to_Disable_and_Bypass_Windows_Management_Instrumentation_LABSCon_Edition/index.html
[25] Microsoft. (2021). NtQuerySystemInformation function (winternl.h). Retrieved from https://learn.microsoft.com/en-us/windows/win32/api/winternl/nf-winternl-ntquerysysteminformation
[26] Microsoft. (2022). Get started with Windows debugging. Retrieved from https://learn.microsoft.com/en-us/windows-hardware/drivers/debugger/getting-started-with-windows-debugging
[27] Microsoft. (2022). Microsoft System Call Table. Retrieved from https://j00ru.vexillium.org/syscalls/nt/64
[28] MITRE. (2021). Exploitation for Privilege Escalation. Retrieved from https://attack.mitre.org/techniques/T1068
[29] Odzhan. (2019). Windows Process Injection: Print Spooler. Wordpress. Retrieved from https://modexp.wordpress.com/2019/03/07/process-injection-print-spooler
[30] Oxcsandker. (2022). Offensive Windows IPC Internals 3: ALPC. Retrieved from https://csandker.io/2022/05/24/Offensive-Windows-IPC-3-ALPC.html
[31] Pei, K. Gu, Z., Saltaformaggio, B., Ma, S., Wang, F., Zhang, Z., Si, L., Zhang, X., Xu, D. (2016). HERCULE: Attack story reconstruction via community discovery on correlated log graph 2016. Purdue University, IBM T.J. Watson Research Center. Retrieved from https://www.cs.purdue.edu/homes/dxu/pubs/HERCULE.pdf
[32] Pogonin, D., Korkin I.. (2022). Microsoft Defender willbe defended: MemoryRanger prevents blinding Windows AV. The 15th Annual ADFSL Conference on Digital

Forensics, Security and Law, 2022. Retrieved from https://commons.erau.edu/cgi/viewcontent.cgi?article=1472&context=adfsl

[33] Poslušný, M. (2022). Signed kernel drivers – Unguarded gateway to Windows' core. ESET. Retrieved from https://www.welivesecurity.com/2022/01/11/signed-kernel-drivers-unguarded-gateway-windows-core

[34] Russinovich, M. (2022). WinObj v3.14. Retrieved from https://learn.microsoft.com/en-us/sysinternals/downloads/winobj

[35] Russinovich, M., Johnson, K. (2020, 04 28). LiveKd v5.63. Retrieved from https://learn.microsoft.com/en-us/sysinternals/downloads/livekd

[36] Sanseo. (06 02 2023). Sliver Malware With BYOVD Distributed Through Sunlogin Vulnerability Exploitations. ASEC. https://asec.ahnlab.com/en/47088/

[37] SecureAuth. (2020). GIGABYTE Drivers Elevation of Privilege Vulnerabilities. Retrieved from https://www.secureauth.com/labs-old/gigabyte-drivers-elevation-of-privilege-vulnerabilities

[38] Statcounter. (2023, January 09). Retrieved from https://gs.statcounter.com/os-market-share/desktop/worldwide

[39] Teodorescu, C., Golchikov, A., Korkin I. (2022a). Binarly. Blasting Event-Driven Cornucopia - WMI Edition. Black Hat 2022. Retrieved from https://binarly.io/posts/Black_Hat_2022_Blasting_Event_Driven_Cornucopia_WMI_edition/index.html

[40] Teodorescu, C., Golchikov, A., Korkin I. (2022b). Binarly. LABScon 2022: Attack on WMI Client. Binarly Research. Retrieved from https://www.youtube.com/watch?v=hUc4HmQTdUI

[41] Teodorescu, C., Golchikov, A., Korkin I. (2022c). Binarly. LABScon 2022: Attack on WMI Service. Binarly research. Retrieved from https://www.youtube.com/watch?v=41dew13Tr9A

[42] Teodorescu, C., Korkin, I. (2022d). Blinding Endpoint Security Solutions: WMI Attack Vectors. Ekoparty 2022. Retrieved from https://binarly.io/events/Blinding_Endpoint_Security_Solutions_WMI_attack_vectors/index.html

[43] Voronovitch, E. (2022). New Milestones for Deep Panda: Log4Shell and Digitally Signed Fire Chili Rootkits. Retrieved from https://www.fortinet.com/blog/threat-research/deep-panda-log4shell- fire-chili-rootkits